\documentclass[12pt,draft]{elsart}%
\usepackage{amsmath}%
\usepackage{amsfonts}%
\usepackage{amssymb}%

\begin{document}

\begin{frontmatter}


\title{Spin waves in the magnetized plasma of a supernova and its excitation by
neutrino fluxes }


\author{V.N. Oraevsky, V.B. Semikoz, A.S. Volokitin }

\address{The Institute of Terrestrial Magnetism, Ionosphere and
\\Radio Wave Propagation of the Russian Academy of Sciences,\\IZMIRAN, Troitsk,
Moscow Region,142190,Russia.\\}

\begin{abstract}

The spin effects on electromagnetic waves in a strongly magnetized
plasma with rare collisions is considered with the help of
relativistic kinetic equations, which takes into account the
electron spin dynamics in the selfconsistent electric and magnetic
fields. It is shown that for electromagnetic waves propagating
almost perpendicular to ambient magnetic field the spin effects
become essential in the vicinity of electron gyrofrequency and the
corresponding wave dispersion and growth rate of the
electromagnetic spin waves in the presence of intense quasi
monoenergetic fluxes of neutrino is determined.

\end{abstract}


\end{frontmatter}


We consider the spin wave propagation in a dense plasma with the strong
magnetic field $\mathbf{B}_{0}=\left(  0,0,B_{0}\right)  $ and find in
the case of the quasi-perpendicular wave propagation in the cold
magnetized electron gas, $k_{z}\ll k_{\perp}$, a new eigen mode with
the spectrum given by the paramagnetic spin resonance of electrons,
$\omega$ \ $\approx\Omega _{e}$. The excitation of spin waves in solids
(ferromagnets) by the electron beam is well-known in literature
\cite{Akhiezer}. It should be kept in mind that we consider here the
Fermi gas of free electrons in contrast to the quasi-particle approach
in a condensed matter and neglect also the exchange interaction of
electrons since the long-range forces are dominant in a plasma. The
suitable object for the appearance of spin waves in plasma would be a
polarized electron gas of a magnetized supernova (SN). The powerful
neutrino flux can excite spin waves there analogously to the possible
excitation of plasma waves in an isotropic SN plasma \cite{Bingham}.

Let us derive the spectrum of spin waves in a magnetized plasma. The system of
the self-consistent Relativistic Kinetic Equations (RKE) for the electron
number density distribution $f^{\left(  e\right)  }=f^{\left(  e\right)
}\left(  \mathbf{p},\mathbf{x},t\right)  $ and the electron spin density
distribution $\mathbf{S}^{(e)}\mathbf{=S}^{\left(  e\right)  }\left(
\mathbf{p},\mathbf{x},t\right)  $, has the form%

\begin{equation}
\left(  \frac{\partial}{\partial t}+\left(  \mathbf{{v}\nabla}\right)
\right)  f^{\left(  e\right)  }=e\left(  \mathbf{{E}+\frac{1}{c}\left[
{v,B}\right]  }\right)  \frac{\partial f^{\left(  e\right)  }}{\partial
\mathbf{p}}+St\left(  f^{\left(  e\right)  }\right)  ,\label{Vlasov}%
\end{equation}%
\[
\left(  \frac{\partial}{\partial t}+\left(  \mathbf{{v}\nabla}\right)
\right)  \mathbf{{S}^{(e)}{=}\left[  e\left(  {E}+\frac{1}{c}\left[
{v,B}\right]  \right)  \frac{\partial}{\partial{p}}\right]  {S}^{(e)}+ }
\]
$\ $%
\begin{equation}
+\frac{2\mu_{B}}{\hbar}\left[  \frac{\left[  \mathbf{{S}^{(e)},{B}}\right]
}{\gamma}+\frac{\mathbf{{E}\left(  {vS}^{(e)}\right)  -{v}\left(  {ES}%
^{(e)}\right)  }}{\left(  1+\gamma\right)  c}\right]  +St\left(
\mathbf{{S}^{(e)}}\right), \label{Spin1}%
\end{equation}
For our purpose the exact forms of collisional terms are not essential, or
considering small perturbations of distribution functions with respect to
equilibrium ones, it is possible to present them as%
\begin{equation}
St\left(  f\right)  = - \nu_{e}\left(  f-f_{0}\right)  ,\;\;St\left(
\mathbf{{S}^{(e)}}\right)  = - \nu_{s}\left(  \mathbf{{S}^{(e)}-{S}_{0}^{(e)}%
}\right) \label{Stoss}%
\end{equation}
with some effective collisional frequencies $\nu_{e}$, $\nu_{s}$.

Equations (\ref{Vlasov}) and (\ref{Spin1}) are completed by the Maxwell equations%

\begin{equation}
\frac{\partial\mathbf{B}}{c\partial t}=\left[  \nabla,\mathbf{E}\right]
,\;c\left[  \nabla,\mathbf{B}\right]  +\frac{\partial\mathbf{E}}{\partial
t}=4\pi\mathbf{j,}\label{Maxwell}%
\end{equation}
where the total electron current $\mathbf{j(x},t\mathbf{)=j}^{conv}%
\mathbf{(x},t\mathbf{)}+\mathbf{j}^{mag}\mathbf{(x},t\mathbf{)}$ \ consists of
the convection current of electrons \ $\mathbf{j}^{conv}\mathbf{(x}%
,t\mathbf{)=-}e\int d^{3}p\mathbf{v}f^{\left(  e\right)  }\left(
\mathbf{p},\mathbf{x},t\right)  $, and the magnetization current \cite{Oraevsky01}%

\begin{equation}
\mathbf{j}^{mag}(\mathbf{x},t)\mathbf{=}\mu_{B}\int\frac{d^{3}p_{e}}{\left(
2\pi\right)  ^{3}}\left(  \frac{\left[  \mathbf{\nabla,S}^{\left(  e\right)
}\right]  }{\gamma}-\frac{\left(  \mathbf{v\nabla}\right)  \left[
\mathbf{v},\mathbf{S}^{\left(  e\right)  }\right]  /c^{2}}{1+\gamma}\right)
.\label{magcurrent}%
\end{equation}
Here $e=\mid e\mid$ is the electric (proton) charge; $\gamma=\varepsilon
_{p}/m_{e}c^{2}$is the electron gamma-factor; $\mu_{B}=e\hbar/2m_{e}%
c=5.79\times10^{-5}~eV~T^{-1}$ is the Bohr magneton; $\mathbf{B}$%
\textbf{\ }$\rightarrow\mathbf{B}_{0}+\mathbf{B}$ is the total
magnetic field.

Note that only in the non-relativistic (NR)\ limit $v/c\rightarrow0$,
$\gamma=1$,\ the magnetization current takes the usual form $\mathbf{j}%
^{mag}(\mathbf{x},t)=c\left[  \mathbf{\nabla,M(x,}t\mathbf{)}\right]  $, where
the magnetization $\mathbf{M}\left(  \mathbf{x},t\right)  $ is given by%
\begin{equation}
\mathbf{M(x},t\mathbf{)}=\mu_{B}\int\frac{d^{3}p}{\left(  2\pi\right)  ^{3}%
}\mathbf{S}^{\left(  e\right)  }\left(  \mathbf{p},\mathbf{x},t\right)
\label{Magnetization}%
\end{equation}
For the thermodynamical equilibrium in the external magnetic field
$\mathbf{B}_{0}$ there appears the mean spin polarization%
\begin{equation}
\mathbf{S}^{\left(  e\right)  }(\varepsilon_{p})=\mathbf{S}_{0}^{\left(
e\right)  }(\varepsilon_{p})=\mathbf{-}\frac{\mu_{B}}{\gamma}\frac{df_{0}%
^{(e)}(\varepsilon_{p})}{d\varepsilon_{p}}\mathbf{B}_{0}\label{meanspin}%
\end{equation}
and the corresponding mean magnetization $\mathbf{M}=\mathbf{M}_{0}=\chi
_{0}\mathbf{B}_{0},$ where the Fermi distribution $\ f_{0}\left(
\varepsilon_{p}\right)  \ $and the electron density \ $n_{e}$\ are given by
\[
f_{0}^{(e)}\left(  \varepsilon_{p}\right)  \mathbf{=}\frac{2}{\hbar^{3}%
}\frac{1}{\exp\left[  \left(  \varepsilon_{e}\left(  p_{e}\right)
-\varepsilon_{F}\right)  /T_{e}\right]  +1},\;n_{e0}=\int\frac{d^{2}p_{e}%
}{\left(  2\pi\right)  ^{3}}f_{0}^{(e)}\left(  \varepsilon_{p}\right)  ,
\]
correspondingly, and the static susceptibility of the polarized
non-relativistic (NR) degenerate electron gas%

\begin{equation}
\chi_{0}=-\mu_{B}^{2}\int\frac{d^{3}p_{e}}{\left(  2\pi\right)  ^{3}%
}\frac{df_{0}\left(  \varepsilon_{p}\right)  }{d\varepsilon_{p}}%
\label{susceptib}%
\end{equation}
is small, $\chi_{0}=\alpha v_{F_{e}}/4\pi^{2}c\ll1$. This is the reason why
the static magnetic induction $B_{0}=\left(  1+4\pi\chi_{0}\right)  H_{0} $
and the magnetic field strength $H_{0}$ practically coincide there.

Taking into account a small value of the static susceptibility for NR plasma
(\ref{susceptib}) one can expect that the electron spin influence the
electromagnetic waves should be important near the electron gyrofrequency
$\Omega_{e}$ only, where such influence takes the resonance character. On the
other hand, the electromagnetic field itself influences the electron
trajectory in the resonant way near the gyrofrequency $\Omega_{e}$, and, in
general, this influence is more stronger than the electron spin-magnetic resonance.

However, there is an exception. The ordinary electromagnetic wave which
propagates strictly across the magnetic field, $\mathbf{k}=\left(
k_{x},0,0\right)  $, $k_{x}=k_{\bot}$, has the linear polarization $E=\left(
0,0,E_{z}\right)  $, $B=\left(  0,B_{y},0\right)  $ due to which the
perturbation of the Larmour rotation of an electron is minimized and one can
expect an appearance of the spin effects. Really, let us consider the case of
the transversal wave propagation in NR magnetized plasma separating the mean
polarization (\ref{meanspin}) and perturbations of electromagnetic fields
$\mathbf{E}$, $\mathbf{B}$,%

\[
\mathbf{S}^{\left(  e\right)  }(\mathbf{p,x},t)=\mathbf{S}_{0}^{\left(
e\right)  }(\varepsilon_{p})+\delta\mathbf{S}^{(e)}\left(  \mathbf{p,x}%
,t\right)  \mathbf{,\;B\rightarrow B}_{0}+\mathbf{B.}
\]

In the Fourier representation, $\frac{\partial}{\partial t}\rightarrow
-i\omega$, $\nabla\rightarrow i\mathbf{k,}$ from the Maxwell equation
(\ref{Maxwell}) rewritten as%
\begin{equation}
\left(  \frac{k_{\bot}^{2}c^{2}}{\omega^{2}}-\varepsilon_{zz}\right)
E_{z}=4\pi\frac{ck_{\bot}}{\omega}\delta M_{y},\label{em-spin0}%
\end{equation}
after the substitution the magnetization component $\delta M_{y}$ given as it
follows from (\ref{Spin1})\ \ by
\[
\delta M_{y}=\mu_{B}\int\frac{d^{3}p}{\left(  2\pi\right)  ^{3}}\frac{\left(
\delta S_{+}^{(e)}(\mathbf{p,k},\omega)-\delta S_{-}^{(e)}(\mathbf{p,k}%
,\omega)\right)  }{2i}=
\]%
\[
=-\mu_{B}^{2}\frac{ck_{\bot}}{\omega}E_{z}\left(  \frac{\Omega_{e}}%
{\omega+i\nu_{s}-\Omega_{e}}-\frac{\Omega_{e}}{\omega+i\nu_{s}+\Omega_{e}%
}\right)  \int\frac{d^{3}p}{\left(  2\pi\right)  ^{3}}\frac{df_{0}%
^{(e)}(\varepsilon_{p})}{d\varepsilon_{p}},
\]
we obtain the dispersion equation for the cold magnetized plasma%
\begin{equation}
\frac{k_{\bot}^{2}c^{2}}{\omega^{2}}-1+\frac{\omega_{p}^{2}}{\omega\left(
\omega+i\nu_{e}\right)  }=2\pi\mu_{B}^{2}\frac{c^{2}k_{\bot}^{2}n_{e0}}%
{\omega^{2}\varepsilon_{F}}\frac{\Omega_{e}}{\omega-\Omega_{e}}%
,\label{disperperp}%
\end{equation}
where the r.h.s. represents the electron spin-magnetic resonance at the
gyrofrequency $\omega\sim\Omega_{e}$; $\omega_{p}=\sqrt{4\pi e^{2}n_{e0}%
/m_{e}}$ is the plasma frequency. Obviously, the l.h.s. of this dispersion
equation gives the standard ordinary wave spectrum accounting for the electron
collisions in plasma if one neglects the electron spin in the r.h.s. in
(\ref{disperperp}). However, if the ordinary wave frequency is far from
$\Omega_{e}$, $\omega_{p}^{2}+k_{\perp}^{2}c^{2}\neq\Omega_{e}^{2}$, there
appears the \textit{new branch} : the electron spin wave with the frequency
\begin{equation}
\omega\simeq\Omega_{e}\left(  1+\frac{2\pi\mu_{B}^{2}n_{0}}{\varepsilon_{F}%
}\frac{k_{\perp}^{2}c^{2}}{k_{\perp}^{2}c^{2}+\omega_{p}^{2}\left(
1-\frac{i\nu_{e}}{\omega}\right)  -\Omega_{e}^{2}}\right)  -i\nu
_{s},\label{spinwaveperp}%
\end{equation}
where: (i) for a strong magnetic field one can neglect the electron
collisions, $\nu_{e},\nu_{s}\ll\Omega_{e},$ and (ii) even for a slightly
relativistic dense electron gas, $n_{0}=(p_{F}/\hbar)^{3}/3\pi^{2}%
\preceq\lambda_{e}^{-3}=(m_{e}c/\hbar)^{3},$ $\varepsilon_{F}\sim m_{e}c^{2},$
the correction in parentheses is small, $\sim\alpha$.

Note, that at $\mu_{B}^{2}B_{0}^{2}/\varepsilon_{F}mc^{2}\ll1$ the main
dissipation of the spin mode,
\[
Im~\omega=\nu_{s}+\Omega_{e}\frac{2\pi\mu_{B}^{2}n_{0}}{\varepsilon_{F}%
}\frac{\frac{\nu_{e}}{\Omega_{e}}\omega_{p}^{2}k_{\bot}^{2}c^{2}}{\left(
k_{\perp}^{2}c^{2}+\omega_{p}^{2}-\Omega_{e}^{2}\right)  ^{2}+\left(
\frac{\nu_{e}}{\Omega_{e}}\omega_{p}^{2}\right)  ^{2}}\simeq
\]%
\begin{equation}
\simeq\nu_{s}+\nu_{e}\frac{\omega_{p}^{4}k_{\bot}^{2}c^{2}\frac{\mu_{B}%
^{2}B_{0}^{2}}{\varepsilon_{F}mc^{2}}}{\Omega{}_{e}^{2}\left(  k_{\bot}%
^{2}c^{2}+\omega_{p}^{2}-\Omega{}_{e}^{2}\right)  ^{2}}\label{damps}%
\end{equation}
is due to the collisional relaxation of electron spins $\sim\nu_{s}$.
Special case $k_{\bot}^{2}c^{2}+\omega_{p}^{2}=\Omega{}_{e}^{2}$, which
corresponds to
the coupling of spin and electromagnetic (with $\omega^{2}=k_{\perp}^{2}%
c^{2}+\omega_{p}^{2}$) modes, is not considered here. It is reasonable to
suppose, that $\nu_{s}$ and $\nu_{e}$ are of the same order and in a plasma of
SN shells are determined by Coulomb collisions. Thus, we can estimate the
collision frequency $\nu_{s}$ according to
\begin{equation}
\nu_{s}\sim\nu_{e}\sim n_{e}v_{e}\pi\Bigl(\frac{e^{2}}{\varepsilon_{e}}%
\Bigr)^{2}\ln\Lambda~,\label{nue}%
\end{equation}
where $v_{e}$ is the typical velocity and $\varepsilon_{e}$ is the typical
energy of colliding electrons, $\ln\Lambda\sim10-20$ is the Coulomb logarithm.
For the electron density $n_{e}\geq10^{31}$ cm$^{-3}$ plasma is degenerated
($\varepsilon_{F}\geq m_{e}c^{2}\geq T_{e}$) and an estimation of $\nu_{s}$ in
(\ref{nue}) must be multiplied by the factor $\sim\left(  T_{e}/\varepsilon
_{F}\right)  ^{2}$. In such case putting $v_{e}\sim c$ and $\varepsilon
_{e}\sim\varepsilon_{F}$ one gets
\[
\nu_{e}\sim7.5\times10^{16}\left(  \frac{n_{e}}{10^{31}}\right)  \left(
\frac{m_{e}c^{2}}{\varepsilon_{F}}\right)  ^{2}\left(  \frac{T_{e}%
}{\varepsilon_{F}}\right)  ^{2}\frac{\ln\Lambda}{10}~sec^{-1}
\]
and $\omega_{p}=1.7835\times10^{20}$ sec$^{-1}$, $\nu_{s}/\omega_{p}%
\sim4.2\times10^{-4}$. At lower electron density $n_{e}=10^{29}$ cm$^{-3}$ in
plasma with $T_{e}\sim m_{e}c^{2}\gg\varepsilon_{F}$ in the similar way we
find the next estimation
\[
\nu_{s}\sim7.\,\allowbreak5\times10^{15}\left(  \frac{n_{e}}{10^{29}}\right)
\min\left[  1,\left(  \frac{T_{e}}{m_{e}c^{2}}\right)  ^{1/2}\right]  \left(
\frac{e^{2}}{T_{e}}\right)  ^{2}\frac{\ln\Lambda}{10}\sim10^{16}~sec^{-1}
\]
and $\nu_{s}/\omega_{p}\sim\allowbreak5.\,\allowbreak\,6\times10^{-4}$. In the
more general case of the quasi-perpendicular wave propagation $k_{z}\neq0,$
$k_{z}\ll k_{\bot},$ one can find the bounds on the spin wave existence in a
polarized electron gas. Substituting the electric field component $E_{x}\neq0$
expressed through $E_{z}\neq0$ one obtains from the Maxwell equation
(\ref{Maxwell}) the dispersion equation
\begin{equation}
\frac{k_{\bot}^{2}c^{2}}{\omega^{2}}-\varepsilon_{zz}-\frac{(\frac{c^{2}%
k_{z}k_{\bot}}{\omega^{2}}+\varepsilon_{zx})}{(\frac{k_{\bot}^{2}c^{2}}%
{\omega^{2}}-\varepsilon_{xx})}\frac{k_{\bot}k_{z}c^{2}}{\omega^{2}}%
=4\pi\frac{ck_{\bot}}{\omega}\frac{M_{y}}{E_{z}},\label{dispernew}%
\end{equation}
where in the l.h.s. the permittivity tensor components in the third term (both
$\varepsilon_{zx}$ and $\varepsilon_{xx})$ contain the resonance terms
$\sim(\omega-\Omega_{e})^{-1}$. However, \ in the vicinity $\Omega_{e}$ this
term is the\ correction of the order $\sim(k_{z}/k_{\bot})^{2}\ll1$\ comparing
with the spin resonance term in the r.h.s. since $\varepsilon_{zx}%
\ll\varepsilon_{xx}.$ There appears also the resonance term $\sim
(\omega-\Omega_{e})^{-1}$ in the diagonal component $\varepsilon_{zz}$
connected, in particular, with the cyclotron damping at the Doppler resonance
$k_{z}v_{z}=\omega-n\Omega_{e}$, $n=\pm1,....$ Combining all these corrections
one finds the new dispersion equation%
\begin{equation}
k_{\bot}^{2}c^{2}+\omega_{p}^{2}-\omega^{2}+\frac{\omega_{p}^{2}\omega
}{\left(  \Omega_{e}-\omega\right)  }A_{1}-\frac{k_{\bot}k_{z}\varepsilon
_{F}A_{1}k_{\bot}k_{z}c^{2}}{2\left(  \omega-\Omega_{e}\right)  m_{e}%
\Omega_{e}A_{2}}=2\pi\mu_{B}^{2}\frac{c^{2}k_{\bot}^{2}n_{0}}{\varepsilon_{F}%
}\frac{\Omega_{e}}{\omega-\Omega_{e}},\label{dispernew1}%
\end{equation}
where additional resonance terms in the l.h.s. while competing with the spin
resonance contribution in the r.h.s. are given by the integrals%
\[
A_{1}=-\frac{1}{n_{0}}\int J_{1}^{2}\left(  k_{\bot}\rho_{L}\right)
m_{e}v_{z}^{2}dp_{z}d\pi p_{\bot}^{2}\frac{df}{d\varepsilon},
\]%
\[
A_{2}=-\frac{\varepsilon_{F}}{n_{0}}\int J_{1}^{2}(k_{\bot}\rho_{L})dp_{z}d\pi
p_{\bot}^{2}\frac{df_{0}^{(e)}(\varepsilon_{p})}{d\varepsilon_{p}}.
\]
If the ordinary wave frequency is far from the electron gyrofrequency
$\Omega_{e}$ , $\omega_{p}^{2}+k_{\bot}^{2}c^{2}\neq\Omega_{e}^{2},$ we find
finally from eq. (\ref{dispernew1}) the spin wave spectrum
\begin{equation}
\omega-\Omega_{e}=\frac{\Omega_{e}c^{2}k_{\bot}^{2}}{k_{\bot}^{2}c^{2}%
+\omega_{p}^{2}-\Omega_{e}^{2}}\left[  2\pi\mu_{B}^{2}\frac{n_{0}}%
{\varepsilon_{F}}+\frac{k_{z}^{2}\varepsilon_{F}A_{1}}{2m_{e}\Omega_{e}%
^{2}A_{2}}-\omega_{p}^{2}\frac{A_{1}}{c^{2}k_{\bot}^{2}}\right]
.\label{finalspectrum}%
\end{equation}
While the second term within brackets is small for the quasi-perpendicular
wave propagation $k_{z}\ll k_{\bot},$ the third one coming from the tensor
component $\varepsilon_{zz}$ would be small too for the strong magnetic fields
obeying the inequality%
\begin{equation}
\frac{p_{F}}{m_{e}c}\ll\frac{\mu_{B}B_{0}}{2\left(  m_{e}c^{2}\right)
}=\frac{1}{4}\left(  \frac{B_{0}}{B_{c}}\right). \label{condition}%
\end{equation}
One can easily see from the last inequality that for a slightly relativistic
electron gas $p_{F}\preceq m_{e}c$\ spin waves exist if the magnetic field
$B_{0}$ is stronger than the Schwinger field \ $B_{c}=4.41\times10^{13}$
\ Gauss$,$ $B_{0}\gg B_{c},$ otherwise, for a moderate field $B_{0}\leq$
$B_{c}$\ the electron density should be deluted in the NR plasma, $p_{F}\ll
m_{e}c$. In the Standard Model of electroweak interactions(SM) the system of
the corresponding RKE's including the neutrino one is derived by the
Bogolyubov method analogously to the case of an isotropic lepton plasma
\cite{Semikoz}. For the case of NR plasma the linearized electron spin RKE
\ (\ref{Spin1}) is completed in SM by the main weak interaction term coming
from the $\nu e-$scattering amplitude \cite{Oraevsky01}
\begin{equation}
\frac{G_{F}\sqrt{2}c_{A}f_{0}^{(e)}(\varepsilon_{p})}{m_{e}}\int\frac{d^{3}%
q}{(2\pi)^{3}}\nabla\delta f^{(\nu)}(\mathbf{q},\mathbf{x}%
,t),\label{additional}%
\end{equation}
where $G_{F}$ is the Fermi constant; $c_{A}=\mp0.5$ is the axial weak coupling
with upper (lower) sign for electron (muon or tau) neutrinos correspondingly.
Let us consider the monoenergetic neutrino beam $f_{0}^{(\nu)}(\mathbf{q}%
)=(2\pi)^{3}n_{\nu0}\delta^{(3)}(\mathbf{q-q}_{0})$, which propagates
in a polarized electron gas across the magnetic field
$\mathbf{B}_{0}=(0,0,B_{0})$ , namely along
$\mathbf{n}_{0}=\mathbf{q}_{0}/q_{0}=\left(  1,0,0\right)  $. Hence we
should find the solution of the linearized neutrino RKE \ for the total
(Lorentz-invariant) \ number density distribution $f^{(\nu
)}(\mathbf{q,x},t)=f_{0}^{(\nu)}(\mathbf{q})+\delta f^{(\nu)}(\mathbf{q,x}%
,t)$,
\begin{equation}
\frac{\partial\delta f^{(\nu)}\left(  \vec{q},\vec{x},t\right)  }{\partial
t}+\vec{n}\frac{\partial\delta f^{(\nu)}\left(  \vec{q},\vec{x},t\right)
}{\partial\vec{x}}+F_{j\mu}^{(V)}(\vec{x},t)\frac{q^{\mu}}{\varepsilon_{q}%
}\frac{\partial f_{0}^{(\nu)}(\vec{q})}{\partial q_{j}}+F_{j\mu}^{(A)}(\vec
{x},t)\frac{q^{\mu}}{\varepsilon_{q}}\frac{\partial f_{0}^{(\nu)}(\vec{q}%
)}{\partial q_{j}}=0,\label{neutrino2}%
\end{equation}
which obeys the neutrino current conservation $\partial j_{\mu}^{(\nu
)}(\mathbf{x},t)/\partial x_{\mu}=0$ automatically since the weak interaction
terms have the Lorentz \ structure and are given by the antisymmetric tensors
$F_{j\mu}^{(V,A)}(\vec{x},t)$,%
\begin{align}
F_{j0}^{(V)}(\vec{x},t)/G_{F}\sqrt{2}c_{V}  & =-\nabla_{j}\delta n^{(e)}%
(\vec{x},t)-\frac{\partial\delta j_{j}^{(e)}(\vec{x},t)}{\partial
t}~,\nonumber\label{tensors}\\
F_{jk}^{(V)}(\vec{x},t)/G_{F}\sqrt{2}c_{V}  & =e_{jkl}(\nabla\times\delta
\vec{j}^{(e)}(\vec{x},t))_{l}~,\nonumber\\
\sqrt{2}F_{j0}^{(A)}(\vec{x},t)/G_{F}c_{A}  & =-\nabla_{j}\delta A_{0}%
^{(e)}(\vec{x},t)-\frac{\partial\delta A_{j}^{(e)}(\vec{x},t)}{\partial
t}~,\nonumber\\
\sqrt{2}F_{jk}^{(A)}(\vec{x},t)/G_{F}c_{A}  & =e_{jkl}(\nabla\times\delta
\vec{A}^{(e)}(\vec{x},t))_{l}~.
\end{align}
Here $j_{\mu}^{(\nu)}(\mathbf{x},t)=\int(q_{\mu}/\varepsilon_{q})f^{(\nu
)}(\mathbf{q,x},t)d^{3}q/(2\pi)^{3}$ and\newline $\delta j_{\mu}%
^{(e)}(\mathbf{x},t)=\int(p_{\mu}/\varepsilon_{p})\delta f^{(e)}%
(\mathbf{p,x},t)d^{3}p/(2\pi)^{3}$ are four-vectors of the neutrino current
density\ and the electron current density perturbation correspondingly;
$\delta A_{\mu}^{(e)}(\mathbf{x},t)=m_{e}\int(d^{3}p/(2\pi)^{3})\delta a_{\mu
}^{(e)}(\mathbf{p,x},t)$ is the axial four-vector of the spin density
perturbation where the axial four-vector $\delta a_{\mu}^{(e)}(\mathbf{p,x}%
,t)$ has the components%
\begin{equation}
\delta a_{\mu}^{(e)}(\mathbf{p,x},t)=\left[  \frac{\mathbf{p}\delta
\mathbf{S}^{(e)}(\mathbf{p,x},t)}{m_{e}};\delta\mathbf{S}^{(e)}(\mathbf{p,x}%
,t)+\frac{\mathbf{p}(\mathbf{p}\delta\mathbf{S}^{(e)}(\mathbf{p,x},t))}%
{m_{e}(\varepsilon_{e}+m_{e})}\right]  ~.\label{Pauli}%
\end{equation}
The latter is the statistical generalization of the Pauli-Luba\'{n}ski
four-vector $a_{\mu}$, \cite{Lifshits}%
\[
a_{\mu}(\mathbf{p})=\left[  \frac{\mathbf{p\varsigma}}{m_{e}}%
;\mathbf{\varsigma}+\frac{\mathbf{p}(\mathbf{p\varsigma})}{m_{e}%
(\varepsilon_{e}+m_{e})}\right]  .
\]
Note that the neutrino RKE (\ref{neutrino2}) differs from the result of
\cite{Bingham} by the last term which is stipulated by the parity violation
through the axial vector currents contribution to weak interactions.
Substituting the solution of the neutrino RKE (\ref{neutrino2}) into eq.
(\ref{additional}) and properly solving the modified spin RKE (\ref{Spin1})
one finds the dispersion equation for spin waves enhanced in NR plasma by the
neutrino beam,
\begin{align}
& \left(  (\omega+i\nu_{s})^{2}-\Omega_{e}^{2})\right)  \left(  \omega
+i\nu_{s}-\frac{c_{A}^{2}\Delta^{(\nu)}}{2}A_{z}^{(\nu)}k_{z}\right)
-\nonumber\label{spectrum}\\
& -\frac{(\omega+i\nu_{s})^{2}c_{A}^{2}\Delta^{(\nu)}}{4}(A_{-}^{(\nu)}%
k_{+}+A_{+}^{(\nu)}k_{-})=0~,
\end{align}
where the spin collision frequency $\nu_{s}$ is estimated in (\ref{nue}); the
dimensionless parameter $\Delta^{(\nu)}=2G_{F}^{2}n_{e0}n_{\nu0}/m_{e}q_{0}$
is given by the mean densities of the electrons and neutrinos $n_{e0}$,
$n_{\nu0}$, and the vector $A_{i}^{(\nu)}$ is given by%
\begin{equation}
A_{l}^{(\nu)}=\int\frac{d^{3}q}{(2\pi)^{3}}\frac{\hat{f}_{0}^{(\nu
)}(\mathbf{q})}{q}\left(  \frac{[(\mathbf{k}\mathbf{n})^{2}-k^{2}]n_{l}%
(q)}{(\omega-\mathbf{k}\mathbf{n})^{2}}+\frac{(\mathbf{k}\mathbf{n}%
)n_{l}-k_{l}}{\omega-\mathbf{k}\mathbf{n}}\right)  .\label{factorA}%
\end{equation}
From the dispersion equation (\ref{spectrum}) we find the increment of the
neutrino driven streaming instability for the spin waves which arises due to
the \v{C}erenkov resonance $\omega=\mathbf{kn}_{0}c+i\delta=\Omega_{e}%
+i\delta$\thinspace\ and has the form%
\begin{equation}
\delta=\delta_{\nu_{s}=0}\left(  \frac{\delta_{\nu_{s}=0}}{\nu_{s}}\right)
^{1/2}~,\label{modified}%
\end{equation}
where the increment in the absence of collisions $\delta_{\nu_{s}=0}$ is given
by \cite{Oraevsky01}%
\begin{equation}
\delta_{\nu_{s}=0}=\Omega_{e}\frac{\sqrt{3}}{4}(\Delta^{(\nu)})^{1/3}(\sqrt
{2}\mid c_{A}\mid\sin\theta_{q_{0}})^{2/3}.\label{increment}%
\end{equation}
Here $\theta_{q_{0}}$ is the angle between the neutrino beam
direction $\mathbf{n}_{0}$ and the wave vector $\mathbf{k}$. We
neglected here the vector current terms ($\thicksim$ $c_{V}$ )
entering the neutrino RKE (\ref{neutrino2}) $\ $which give a
negligible contribution ($\thicksim c_{A}c_{V}$) in the spin RKE
for \ long wave lengths exceeding the Compton one,
$2\pi/k\gg\hbar/m_{e}c,$ that is the reasonable approximation in
NR plasma.

For the mean neutrino energy in a magnetized supernova (SN) $q_{0}\sim10$MeV
and for the NR plasma outside the neutrinosphere and behind the shock with the
densities $n_{e0}\simeq10^{29}$cm$^{-3},$$n_{\nu0}\simeq10^{32}$cm$^{-3}$the
parameter is very small, $\Delta^{(\nu)}\sim10^{-25},$while the increment
(\ref{increment}) would be big enough, e.g. for the strong magnetic field
$B_{0}=10^{12}$Gauss and the corresponding gyrofrequency $\Omega_{e}%
=1.7\times10^{7}B_{0}=1.7\times10^{19}$ sec$^{-1}$ it reaches the value
$\delta_{\nu_{s}=0}\sim10^{10}$ sec$^{-1}$. One can see that the spin wave
amplitude increases faster than a neutrino passes throw SN envelope
($\sim10^{-3}$sec). However, the value $\delta_{\nu_{s}=0}$ is less than the
spin wave damping due to collisions $\sim\nu_{s}\sim10^{16}~ $sec$^{-1}$
resulting in a decrease of the neutrino driven instability growth rate,
$\sim\delta_{\nu_{s}=0}(\delta_{\nu_{s}=0}/\nu_{s})^{1/2}\sim10^{7}~$%
sec$^{-1}$.

Nevertheless, the energy exchange between plasma and neutrino fluxes by their
interaction with spin waves probably could lead to the shock revival during
the neutrino burst ($\sim10-20$ sec) through the heating of the surrounding
plasma. Really these spin waves are generally coupled to the magnetosonic ones
analogously to the spin waves in ferromagnets \cite{Akhiezer}, or their energy
can be transferred to the electromagnetic and plasma waves at the cross of spectra.

Note that in a strong magnetic field, $\Omega_{e}\succeq$$\omega_{pe},$the
increment (\ref{increment}) is less suppressed for the small angles $
\theta_{q_{0}}\leq\arccos\left(  <v>/c\right)  $ than the corresponding one
for plasma waves\ in the isotropic plasma \cite{Bingham},
\[
\delta_{\nu_{s}=0} =\omega_{p}\frac{\sqrt{3}}{2}(\Delta^{(\nu)})^{1/3}\left(
\sqrt{2}c_{V} \frac{\sin^{3}\theta_{q_{0}}}{\cos^{2}\theta_{q_{0}}}\right)
^{2/3},
\]
that allows to decrease the inevitable Landau damping of collective modes
excited by the neutrino beam when spin waves propagate through a relativistic
plasma with the mean electron velocity $<v>\sim c$.

There is the second advantage of spin waves enhanced via the weak axial vector
currents ($\sim c_{A}=\mp0.5$) instead of \ the case of plasma waves excited
via the weak vector currents $\ $with the small vector coupling in the case of
muon and tau neutrinos (choosing the lower sign in $c_{V}=2\xi\pm0.5$where
$\xi\simeq0.23$is the Weinberg parameter )$.$ This is the reason why authors
\cite{Bingham} considered the case of electron neutrinos only and put for them
$c_{V}\longrightarrow1$. Note that during the main neutrino burst in SN all
neutrino species are produced in the hot SN\ core via the pair annihilation
$e^{+}e^{-}\rightarrow\nu_{a}\tilde{\nu}_{a},$$a=e,\mu,\tau$.

Thus, we conclude that spin waves (\ref{finalspectrum}) exist in a magnetized
plasma with the strong magnetic field obeying the condition (\ref{condition})
and can be efficiently excited, e.g. in a magnetized SN by a powerful neutrino
flux that could be an effective collective mechanism to revive the shock with
the following blust of the SN envelope.

This work was partially supported for A.V. and V.S. by the RFBR grant 00-02-16271.

\end{document}